# 1 Exoplanet Chemistry

*Katharina Lodders*

Planetary Chemistry Laboratory, Dept. Of Earth & Planetary Sciences and McDonnell Center for the Space Sciences, Washington University, One Brookings Drive, Saint Louis, MO 63130

Abstract: The characteristic chemistry of terrestrial planets and, in particular, of giant planets rich and poor in He and $H_2$ are described.

## 1.1 A Goodly Gallery of Planets

The current state of observations on the chemical nature of exoplanets is comparable to that of the early days of solar system planetary science, when not much more than the orbits and densities of planets were known. Information about exoplanet chemical compositions from direct observations is still limited. However, chemical and physical observations and models for planets in our solar system developed since the 1950s should provide the tools to understand other planets, if models for solar system planets are indeed correct. As such, the observational data to come in the following years undoubtedly will test the fine-tuned planetary formation and evolution models that seem to work well for understanding the planets in our solar system.

There are four broad basic types in our gallery of planets in the solar system and very likely in most of the other planetary systems. The reason for this is the first order similarity of the relative elemental abundances of most stars in the galactic disk and the compounds that the elements can form. Planets originate from the materials in a stellar accretion disk that had a similar composition as the final star itself. This limits the possible chemical nature of the building blocks of planets, despite the wide range of possible physical properties in the different accretion disks around other stars. Different condensed phases in an accretion disk are stable to different temperature limits; refractory phases are stable up to high temperatures and volatile phases are only stable at relatively low temperatures. Thus, the formation or thermal destruction of solids at different temperatures can lead to a fractionation of refractory elements in the solids from the more volatile elements that are in the gas.

Figure 1.1 shows how the four types can be categorized by their chemical nature. Terrestrial-like planets, composed mainly of rocky substances defined below, are represented in our solar system by Mercury, Venus, Earth, Mars, Ceres and, although not planets, by a large number of asteroids. Terrestrial planets can have atmospheres which are, however, insignificant to the total planetary mass budget. The second type is planets largely made of rocky and icy materials (e.g., $H_2O$, $CH_4$, $N_2$ ices), which may have tenuous atmospheres composed of icy evaporates. In the solar system, such "Plutonian" planets, as one could call them, are only represented by IAU-defined dwarf planets such as Pluto, Eris, Haumea and Makemake, but most objects beyond Neptune should qualify for this chemical distinction. Some of the asteroids in the outer asteroid belt may also belong to the rocky-icy objects. Giant (Neptunian) planets consist mainly of compounds derived from rocky and icy substances in the accretion disk and contributions from He and $H_2$ to their total masses are considerable but remain below 50%. Uranus and Neptune could only accrete about 10-20% of their total mass as H and He gas, which makes them "gas-poor" giant planets here. The fourth type, the Jovian gas-rich giant planets in which H and He dominate, is represented by Jupiter and Saturn, and they have the closest similarity in elemental composition to their host star, the Sun.

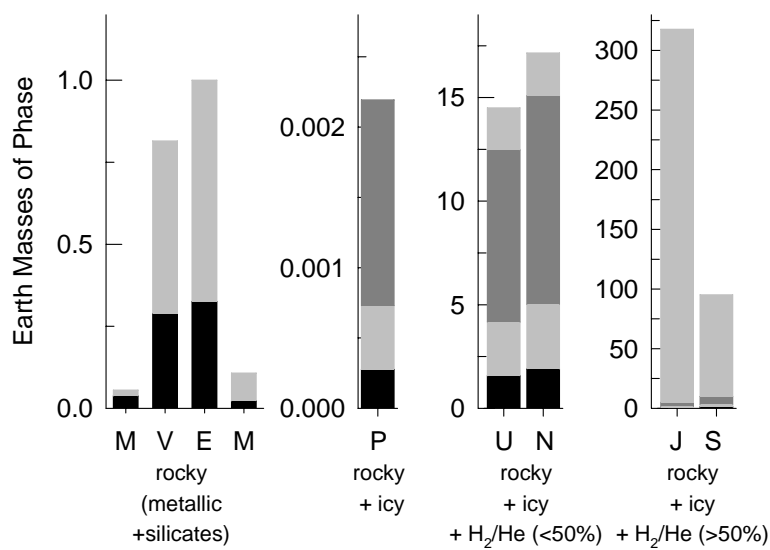

Figure **1.1**. The four chemical types of planets represented by solar system objects. The shading indicates the contributions (in Earth masses) of different phases black=metallic, light grey=silicate, dark grey= various ices, second light grey on top of ices is $H_2$+He gas. From left to right: Rocky (terrestrial) planets mainly consisting of silicates and metallic FeNi(S). Rock plus

ice (Plutonian) planets primarily consisting of rocky and icy solids. Giant (Neptunian) planets consist mainly of rocky and icy substances and less than half is He and $H_2$. Gas-giant (Jovian) planets are dominated in mass by He and $H_2$. Data from Lodders & Fegley 1998.

Silicates and metallic phases define the rocky portion of a planet, however, their proportions are only reasonably well known for the terrestrial planets. For all other objects in Figure 1.1, a silicate/metallic mass ratio according to solar elemental abundances and geochemical considerations was applied to the "rocky" portion. For the giant planets, similar considerations were used to divide the heavy element fractions into "metallic, silicate, and icy" components.

Most known exoplanets have masses between that of Saturn and ten times that of Jupiter (1 $M_{jup}$ = ~318 Earth masses, $M_E$); but smaller Neptune size objects (~0.05 $M_{Jup}$ = ~17 $M_E$), and objects of a few Earth masses (1 $M_E$ = 0.003 $M_{Jup}$) are also known (see the frequent exoplanets updates on Jean Schneider's webpage at http://exoplanet.eu). A few companions to stars reach masses above ~13 $M_{Jup}$, where theoretical models predict that interior deuterium burning is possible. This mass limit is usually taken to divide "planets" from "brown dwarfs" that are also observed to orbit stars. However, this definition may be arbitrary. To make this point, consider the object HD41004B b, which orbits its early M dwarf host star at 0.0177 AU with an orbital period of ~ 1.3 days. This is one of the closest orbits of a non-stellar object around a star currently known. However, HD41004B b has a mass of 18.4 $M_{Jup}$ – clearly above the D-burning mass limit, and is not really counted as a planet.

The atmospheric gas and cloud chemistry in brown dwarfs is not much different from that in gas-giant exoplanets if these objects have similar effective temperatures (see below). The major difference would be the absence of deuterated gases such as $CH_3D$, HDO, $NH_2D$, and HD in brown dwarfs that masquerade as huge planets. If observationally feasible, the presence or absence of D-bearing gases may be a means to distinguish large planets from brown dwarfs, but more importantly, it would finally provide observational measures of the minimum mass required for D-burning.

There is no other practical means to distinguish these objects other than by mass. Differences in possible formation modes could also distinguish these larger objects. Gas giant planets and brown dwarfs can both form by gravitational instabilities in a protostellar accretion disk (Chapter X). Another prominent formation model for giant planets is "core accretion" (Chapter X). This is accretion of larger protocores and subsequent gravitational capture of large quantities of gas (see below), but this mechanism may not work for making massive brown dwarfs in a timely fashion in a short-lived accretion disk. In principle, the different modes of formation can imprint different chemical compositions onto a gas giant planet (see below). However, even the interpretation of the observed chemistry of the well studied gas-giant planets Jupiter and Saturn is not unique. Currently a large number of researchers seem to favor the core accretion model for Jupiter's and Saturn's origins, but the case is not clear-cut. Hence, the much more limited chemical information that is available so far for gas giant planets outside the solar system can not be used to distinguish planets and brown dwarfs by their formation mode.

## 1.2  Elemental ingredients of planets

The densities of exoplanets can be found if their radii and masses can be determined. Radial velocity measurements provide masses (or upper mass limits M sin$i$) for exoplanets, and radius determinations are available for planets observable during transits of their host star (see Chapter 1). Once planet densities are available, planetary overall compositions may be estimated from the chemical and physical properties of plausible components. Several models for gas giant exoplanets evolution, their interiors, and spectra have been described by e.g., Burrows et al. 2000, 2003, 2004, Charbonneau et al. 2007, Chabrier et al. 2007, Fortney et al. 2004, 2008a,b, Guillot, 2008, Guillot et al. 2004, 2006 , Sudarsky et al. 2003. A wide range of mass-radius relations for terrestrial to gas giant planets is discussed by Fortney et al 2007a There are also several different structure types of terrestrial-like planets that have been modeled, based on different components that may result from solar composition element mixtures e.g., Fortney et al. 2007, Seager et al. 2007.

Ultimately, it is the abundance of the chemical elements and the physical properties of their compounds that govern the constitution of a planet. As already mentioned most stars in the solar neighborhood have similar relative elemental abundances as the Sun. Hence, the compounds and their relative amounts available in the accretion disks around other stars should be similar to what was present in the solar planetary accretion disk (i.e., the solar nebula). There are, of course, limitations to this first-order approach because the *distribution* of these compounds in a disk is expected to vary as physical disk properties vary. Planets not only develop around forming G2V dwarf stars like the Sun, but are also found around lower mass F, K and M dwarfs, and around stars with up to ~ 4.5 solar masses that have evolved into G and K giants. Detections of planets around very massive and hotter dwarf stars are not yet reported. Planet formation around early type (OBA) dwarfs may be difficult to achieve as estimated planetary formation and evolution times of a few to tens of million years may coincide with stellar lifetimes, and early mass loss from massive stars may interfere with planet formation in an accretion disk.

The estimated minimum mass of the solar nebula is about 0.01 to 0.02 $M_{sun}$, depending on the solar abundance compilation used and the adopted heavy element fractions for the giant planets. Accretion rates through the disk onto the young Sun are estimated as ~ $10^{-5}$ to $10^{-7}$ $M_{sun}$/year, depending on the accretion stage. However, accretion disks around other stars may have had different total masses than the solar nebula, different accretion rates from their parental molecular clouds, and thus different thermal gradients and evolution time scales of their disks. Disk lifetimes are important because the formation of gas-giant planets requires that $H_2$ and He rich gas is accreted before the disk dissipates. Meteoritic evidence from short-lived radioactive nuclides and observations of accretion disks in young stellar system suggest that disk lifetimes are at least 2-3 Ma, with an upper limit of about 10Ma, which is then the maximum time available for gas-giant planet formation. On the other hand, the disk lifetimes and the presence of nebula gas may not be as critical for the formation of terrestrial type planets which can form as long as rocky planetesimals are abundantly available. The formation of the moon by a giant impact occurred after core formation on the Earth was complete, which was within 30 Ma after the oldest known components in meteorites had been processed in solar nebula gas ~4600 Ma ago (e.g., Kleine et al. 2002). The craters on Moon, Mars, and Mercury are witness to the late bombardment period which lasted until ~3900 Ma ago. Clearly, formation of the Moon and late planetary bombardments are the larger finalizing events in terrestrial planet accretion, but their timing is outside any reasonable lifetimes of the solar nebula.

Depending on location of a star in the galactic disk and on timing of star formation, a star and its accretion disk may also have had different metallicity than the Sun (i.e., heavier elements than He have different abundances relative to H). This additional factor influences the outcome of a planet's final constitution. The metallicity of a star is important because it is observed that metal-rich stars are more likely to harbor planets than stars of lower metallicity (e.g., Gonzalez, 2003). Most of the known exoplanets are likely to be gas giants, but one should expect that the occurrence of rocky planets has a similar preference at metal-rich stars. The formation of planets is plausibly fostered by a metal-rich disk because a higher mass density of the "metallic" elements (in the astronomer's sense) provides more rocky and icy building blocks which should increase the probability of planet accretion.

### 1.2.1 Diagnostics from elemental abundance fractionations

The *elemental* composition of a gas giant planet compared to the composition of its host star may reveal more about its mode of formation, and again the relative amounts of the heavy elements are of particular interest here. A gas-giant planet that is essentially identical in elemental composition to its host star is likely to have formed through gravitational instability in the protostellar accretion disk (e.g., Boss 1997, 2001). On the other hand, in the core accretion model for giant planet formation, a solid proto-core of ~5-10$M_E$ has to form first, which is large enough to gravitationally capture surrounding nebular $H_2$ and He gas (Pollack et al. 1996, Hubickyj et al. 2005, Lissauer & Stevenson 2007, see also Chambers, 2003, and chapter by **XX**). This leads to a gas giant planet rich in elements heavier than He and H when compared to the composition of the primary star if formation of the initial core required accumulation of solids from wider regions in the disk but gas was only accreted locally. (A similar result is obtained when incomplete accretion of He and $H_2$ fails to add to the rocky core abundances to get back abundances of the host star.) This model also leads to the notion that all giant planets formed by this mechanism should have "cores", or more precisely, the equivalent mass of heavy elements distributed somewhere in their interior (e.g. Saumon & Guillot 2004)

In addition to various accretion scenarios that separate solid components from the He/$H_2$ gas, chemical fractionations among heavy elements themselves in the disk can cause non-solar abundance ratios of the heavy elements found in planets. For example, solar composition has a mass ratio of C/Si ~ 3, whereas the entire Earth has a ratio of ~0.0003. This is a consequence of the different volatilities of the chemical compounds formed by C and Si. The most likely place of fractionation of C and Si was in the solar nebula, and the Earth could not accrete much of C- and other volatile element bearing compounds to begin with.

## **1.3** Planetary building blocks

The basic results of chemistry in a H and He rich gas such as the solar nebula remain applicable for systems that evolve with solar or close to solar overall elemental compositions, if temperatures and densities (or total pressures) are not dramatically different. It is well known from condensation calculations that over a wide range of total pressures (i.e., $P < 10^{-3}$ bars), the chemistry in a very $H_2$ and He rich, solar-like gas leads to quite similar minerals and thermal stability sequences (e.g., Grossman & Larimer 1974, Lewis 1974, Lodders 2003 and references therein). For example, usually the most thermally stable solid containing most of the Fe is an FeNi metal alloy; for Mg, it is the Mg-silicate forsterite ($Mg_2SiO_4$), and for sulfur, it is troilite, FeS. Hence, the available planetary building blocks in different disks should be comprised of quite similar oxides, silicates, metals, sulfides, and ices. Aside from the thermal stabilities, the quantities of the building blocks are limited by the abundances of the elements that make them. This simplifies the modeling of possible exoplanet compositions, even if the exact conditions in a former accretion disk and the distribution of the solids are not known.

The radial distribution of the rocky, icy, and gaseous ingredients for making planets is determined by the temperature and total pressure gradients in accretion disks. Therefore even in absence of any possible turbulent redistribution of constituents, the available proportions of the solid planetary building blocks vary with radial location, which will affect the locations, masses, types, and number of resulting planets.

Three principal types of compounds, loosely referred to as rock, ice and gas, are available in accretion disks, and can wind up in planets. When discussing planetary compositions, the terminology rock, ice, and gas is often kept, but it may relate to the nature of the phases that may have accreted to a planet, and not necessarily to the nature of phases that occurs in a planet today. The relative proportions of gas, rock, and ice in the accretion disk determine a planet's possible mass but the equations of state determine the planet's size and which stable phases will be present.

The gas component is mainly $H_2$ and He and includes all the remaining mass that is left after rocky and icy condensates are removed. Depending on temperature and the details of the accretion disk chemistry of ices, inert gases (noble gases, $N_2$) can contribute a tiny mass to the gas. Since H and He make ~98.5% of all mass in overall solar system material, these elements must make the major mass in any gas giant planet that formed by the disk instability mechanism or, in the core accretion model, efficiently captured more than its initial mass of the rocky and icy core (this assumes that there were no changes in the H and He content from post accretion processing of a giant planet that has migrated to close to its host star).

The rocky and icy components (noble gases inclusive) make up ~ 1.5% of the total mass. Figure 1.2 shows the mass percentages in a solar elemental composition system for various rocky and icy substances as a function of temperature. The three panels are for different assumptions of thermodynamic equilibrium, which mainly affect the amount and type of phases present at low temperatures.

The high temperature rock includes elements like Si, Mg, Ca, Al, Ti, etc., that form oxides and silicate minerals. Here FeNi metal alloys and iron sulfide also belong to the "rock" in the environs of planetary accretion disks. Formulas and names of major minerals commonly encountered in planetary science and some mineral systematics are in Table 1. A good working definition of "rock" is a component containing the approximate elemental abundances and consisting of the mineral phases that are observed in undifferentiated meteorites called chondrites. All rocky compounds are relatively refractory and require high temperatures for evaporation or they readily condense out of a gas of solar system composition at relatively high temperatures (see below for condensation). Since rocky material is the last material to evaporate or the first to condense, it is always present among possible solids in an accretion disk. Planets primarily made of such rocky materials - the terrestrial(-like) planets - characterize the chemical make-up, but not necessarily the size of such planets. The solar system abundance of the elements limits the amount of rocky material to ~0.5% of all mass in an accretion disk.

[RockyMinerals]
**Table 1.1** Typical minerals made of more abundant elements

| Mineral group | Endmember mineral | Ideal formula |
|---|---|---|
| Olivine $(Mg,Fe)_2SiO_4$ | Forsterite | $Mg_2SiO_4$ |
| | Fayalite | $Fe_2SiO_4$ |
| Pyroxene $(Mg,Fe,Ca)SiO_3$ | Enstatite | $MgSiO_3$ |
| | Ferrosilite | $FeSiO_3$ |
| | Wollastonite | $CaSiO_3$ |
| Feldspar | Anorthite | $CaAl_2Si_2O_8$ |
| | Albite | $NaAlSi_3O_8$ |
| | Orthoclase | $KalSi_3O_8$ |
| Metal alloys | Iron-nickel | FeNi |
| Sulfides | Troilite | FeS |
| | Pyrrotite | $Fe_{1-x}S$ |
| Oxides | Magnetite | $Fe_3O_4$ |
| Hydrous silicates | Talc | $Mg_3(Si_4O_{10})(OH)_2$ |

At temperatures below ~ 600 K, gas-solid reactions become sluggish and equilibrium may not be reached. As discussed in detail by Fegley (2000), under equilibrium conditions one may expect the formation of sulfide and magnetite from metal, and hydrated silicates from pyroxenes and olivine. However, considering reaction kinetics, only sulfide formation was quantitative within the lifetimes of solar nebula, whereas magnetite formation is a border-line case. The hydration of silicates, usually associated with hydrothermal, high pressure environs, in the "vacuum" of accretion disks can be ruled out since this requires reaction times that are many orders of magnitude above estimated lifetimes of these disks. Therefore, the non-equilibrium cases in Figure 1.2 do not show hydrous silicates as possible planet-building phases.

Icy materials become stable below ~ 200 K, and encompass substances containing C, N, and O. The solids that more or less quantitatively sequester C, N, and O only become stable at low temperatures in low pressure solar composition systems. Hence, these low temperature phases are collectively called "ice". Water ice is the most important ice because O is the third most abundant element in the solar system, and water ice is the most refractory ice among these ices. (Oxygen is also present in rocky materials but the other elements in rock are only abundant enough to bind ~20% of all available solar system oxygen).

In addition to water ice, other possible abundant ices are solid methane, $CH_4$; methane clathrate $CH_4 \cdot nH_2O$; carbon monoxide, $CO$; carbon monoxide clathrate, $CO \cdot nH_2O$; carbon dioxide $CO_2$; carbon dioxide clathrate, $CO_2 \cdot nH_2O$; ammonia, $NH_3$, ammonia monohydrate $NH_3 \cdot H_2O$; molecular nitrogen ice, $N_2$; and nitrogen clathrate, $N_2 \cdot nH_2O$. In the clathrate hydrates, with the generic formula $X \cdot nH_2O$, n = 5-7 is the number of water molecules that characterize a unit structure into which another molecule "X" can be retained as a "guest", which is why clathrates are also called "cage compounds".

One may add solid organic substances to the group of "ices". High-polymer organics can be stable up to temperatures of 400-600K (e.g., Kouchi et al. 2002), which makes them more refractory than water ice.

Gas reactions involving very stable molecules such as CO and $N_2$ are unlikely to reach thermodynamic equilibrium (see Fegley & Prinn 1989, Fegley 2003). This influences the type of low temperature C and N gases, which in turn determines the type and amount of the ices that can form. In addition, the available proportions of the different ices depend on the relative thermal stability of these ices.

The top graph in Figure 1.2 is the equilibrium case where all CO converts to $CH_4$ gas which then condenses as methane clathrate and $CH_4$ ice. Both phases form because the amount of water is insufficient to clathrate all $CH_4$ gas as methane clathrate (solar abundances give C:O about 1:2 but a ratio of 1: (6 to7) is needed for methane clathrate. The thermodynamically stable low temperature gas for nitrogen is ammonia gas, which condenses as ammonia hydrate, $NH_3 \cdot H_2O$.

The formation of methane ice(s) requires that the very stable CO gas, dominant at high temperatures (e.g., >1500 K at $10^{-5}$ bars), was efficiently converted into methane gas at low temperatures, which is unfeasible within typical lifetimes of the disks (see Fegley & Prinn 1989). With inefficient conversion, CO gas remains, and CO clathrates and CO ice condense (Figure 1.2 middle panel). Some CO (several %) is also converted to $CO_2$ since this reaction is kinetically favored and can proceed down to lower temperatures (800-1000K) at low total pressures. Thus, smaller amounts of $CO_2$-bearing ices (equivalent to ~10% of all O) can be present (Fegley & Prinn 1989, Fegley 2003), which, for clarity, are not shown in the non-equilibrium case panels in Figure 1.2. Slow gas phase kinetics also prevents the conversion of $N_2$ to $NH_3$ and $N_2$ gas remains. This leads to $N_2$-clathrate ice (see Fegley & Prinn 1989, Fegley 2003).

Another possibility is that interstellar low-temperature organic substances remained unprocessed or that such substances formed in the outer planetary disks through catalyzed reactions such as in the Fischer-Tropsch-type process, where Fe-metal acts as a catalyst to convert CO and $H_2$ to hydrocarbons. This reduces the amounts of other C-bearing gases and ices.

To first approximation, the exact C, N, and O bearing phases are irrelevant for mass balance as long as all C, N, and non-silicate O are condensed as ices. Condensation of all C, N, and non-silicate O into ices makes about ~1% of all material in a solar composition system. Water ice alone makes up to ~0.6%, depending on whether all remaining O after rock condensation is in $H_2O$ gas or remains in CO gas. However, full retention of the icy phases requires temperatures less than ~ 30 K), and one can expect that the most volatile ices such as CO, $CH_4$, and $N_2$ ice are easily evaporated during formation and evolution of the accretion disk.

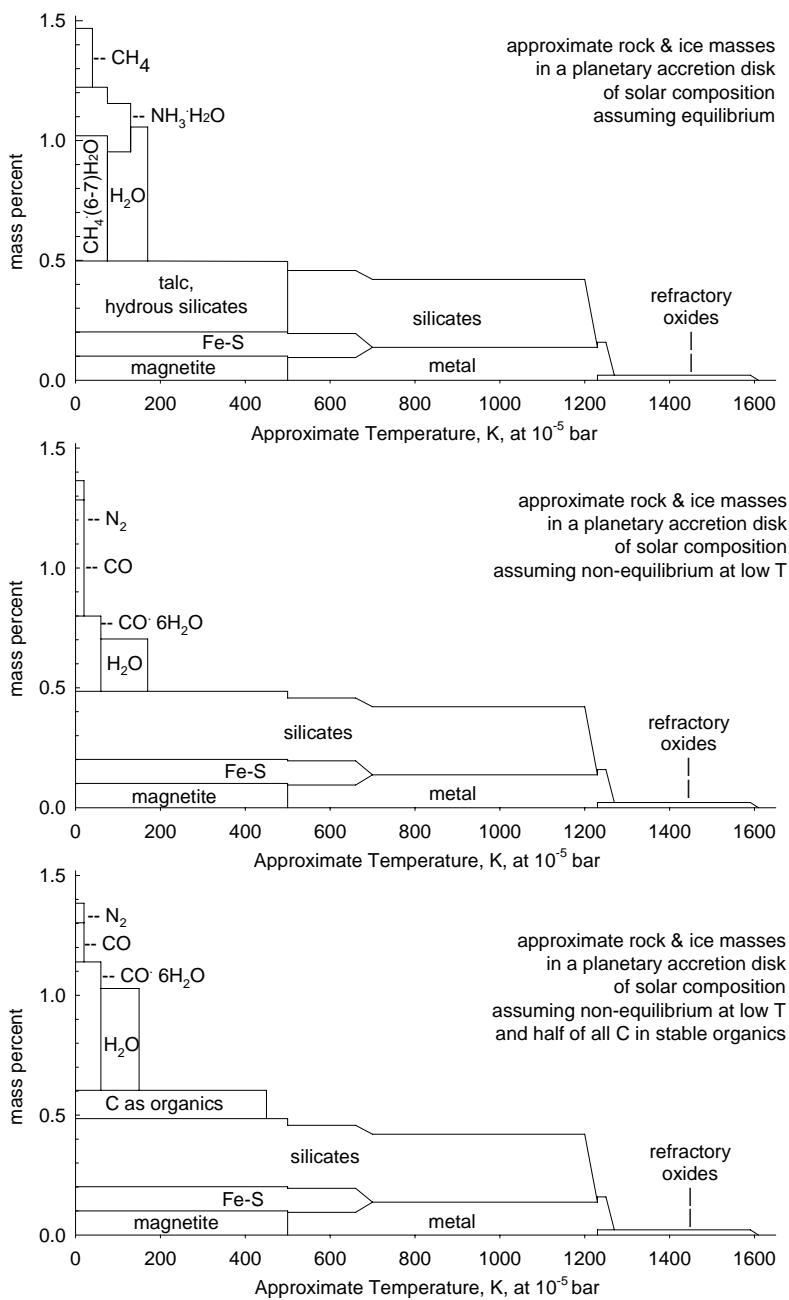

Figure 1.2. The distribution of solid phases as a function of temperature at $10^{-5}$ bar total pressure. The amounts shown are mass% for solar system composition. The top is for the thermodynamic equilibrium case, which is limited by slow gas-gas and gas-solid reaction kinetics at low temperatures. The middle and bottom diagram illustrate possible (and non-unique) low temperature non-equilibrium phases and distributions.

Planets may form by accreting varying amounts of icy and rocky materials but this does not mean that rocks and ices remain in a state of "ice" or "rock" as one typically associates with it. Occasionally it is convenient to collectively equate the amounts of C, N, and O as "ice" in planetary materials which reflects that these elements may have been brought to the planet in low temperature phases. Similarly, "rock" is a convenient summary of all other condensed phases other than "ice". However, temperature and pressure conditions during planetary accretion and subsequent planetary evolution can be quite different from those under which these phases are stable in a low temperature, low pressure accretion disk.

Only small planetary objects such as Kuiper belt objects (=dormant comets?) in the outer solar system are most likely to have preserved their originally accreted icy phases. However, photochemical reactions are known to produce chemical and physical changes in surfaces of "icy" objects. On the other hand, icy phases and rock can thermally decompose during the larger scale differentiation of a planet, or react with other substances once they accreted into larger planetary objects (e.g., reactions of rocky materials with $H_2$ upon accretion into a gas-giant atmosphere).

During accretion of smaller km-size objects to Mars-and Earth size objects, enough gravitational energy is released to allow melting and (partial) evaporation of accreting solids, even if about half the gravitational energy is lost through radiation. The point is that phases present in the accreting bodies may break down, and the elements from such phases react and redistribute into phases that are stable under conditions of the planetary environment. The separation of "rocky" elements into the silicate and metal portions in the Earth and other terrestrial planets is a well known example. Another result of planetary differentiation is the redistribution of rocky elements within the Earth's silicate portion into a crust, the upper mantle of mainly olivine, and a lower mantle containing larger amounts of a high-pressure modification of $(Mg,Fe)SiO_3$ called silicate perovskite (not to be confused with the Ca-Ti mineral perovskite which has a similar structure at normal pressure). By analogy to meteorites, the prime candidates for constituents of the metallic cores in terrestrial planets are the "rocky" elements Fe, Ni and S, which transform into high pressure metal and sulfide phases in planetary interiors, again phases in quite different states than those encountered in the low pressure mineralogy of chondritic meteorites.

Another example for the likely differentiation processes is the accretion of rock and ice to form proto-cores of 5-20 Earth masses, in the core accretion model for Jupiter and other gas giant (exo)planets. These proto-cores may differentiate in a similar manner as the terrestrial planets have differentiated. If heated by gravitational energy release or radioactive decay, icy materials begin to outgas and to form an atmosphere. With estimated formation times of up to ~8 Ma for Jupiter (e.g., Pollack et al. 1996), at least some partial outgassing can be expected before rapid $H_2$ and He gas capture occurs. In the absence of larger amounts of H and He, such an atmosphere should contain the same gases that made the initially accreted ices, adjusted for local temperature and pressure equilibria. Thus, an icy + rocky proto-core as predicted by the core accretion model may show an atmosphere made of $H_2O$, CO, $CO_2$, $N_2$ and possibly some traces of methane and ammonia before evolving into a gas giant planet. If so, hot differentiated and outgassed protocores may be detectable since these gases have characteristic absorption bands.

In the disk instability model, an extended proto-planet (~few 100 Jupiter radii) may develop a rocky core. Recently, Helled & Schubert (2008) addressed the survival of grains and the sedimentation efficiency in the protoplanetary atmospheres of gas-giant planets that form through disk instability. They find that protoplanets of 5 Jupiter masses or larger cannot form silicate cores because temperatures become too high. Smaller proto-planets allow grain settling and silicate "core" formation if contraction rates are low and low internal temperatures (<1300 K) are maintained. However, the temperatures reached during accretion are still too high to allow the accretion of ices to the protoplanet. Thus, the proto-core formation process fractionates refractory rock from more volatile elements (C,N,O) that form ices.

There is, however, no evidence that such "cores" formed during disk instability or in the core accretion model would remain intact as cores in giant planets, presumably still composed of the heavy elements originally present in the initial core structures. The destruction of rocky and icy proto-cores during the gradual growth of a giant planet cannot be excluded. On the other hand, the interior structures (moments of inertia) and thermal evolution models of Jupiter and Saturn require that denser cores are present today (Wuchterl et al. 2000, Saumon & Guillot 2004). The amount of heavy elements in Jupiter may be up to 42 $M_E$, depending on the mass distribution of these elements between the gaseous envelope and the core. Jupiter's interior structure can be consistently modeled without a heavy element core but also with a core of up to $10M_E$. Saturn's structure seem to require core masses made of heavy elements between 10-25 $M_E$, where up to $10M_E$ of heavy elements may be located in the envelope. One larger uncertainty in these interior models is that the equations of state for H and He at the extreme temperatures and pressures within Saturn, Jupiter, and the more massive giant exoplanets are not yet well known. Thus, whether cores are remnants of the original heavy element proto-cores that never eroded, or are results of interior phase separations is not fully clear (see also Stevenson 1985, Guillot et al. 2004).

The destruction of rock and ice certainly happens during the later stages of accretion when smaller planetesimals plunge into a hot and $H_2$-rich atmosphere of a more mature gas giant planet. For example, accreting water ice evaporates as the growing planet is gravitationally heated. Then water vapor becomes a more prominent constituent of the gaseous atmosphere. Another example are solid organic (C-bearing) compounds that react with $H_2$ to form methane, a major gas in cool gas giant planets and also found in gas giant exoplanets. Similarly, iron sulfide contained in rocky accreting material can easily break down and react with $H_2$ to $H_2S$ at temperatures above 600 K. Hydrogen sulfide remains as a gas in a hot giant planet atmosphere (below the $NH_4SH$ cloud base at ~200 K) although it may also be photolyzed by UV light ($\lambda <$ 280 nm) at higher altitudes.

Overall, the content of heavy elements in the observable atmospheres of gas-giant planets is from three principal sources:

(1) Elements present in gaseous form in the planetary accretion disk. In case of the core-accretion model, the stage of gravitational $H_2$ and He capture may also include volatile elements in non-condensed phases CO, $N_2$, $CH_4$, etc, depending on temperature in the protoplanetary accretion disk. In the disk instability model of giant planet formation, all elements accrete in the same proportions from the accretion disk as present in the host star.

(2) If giant planets form through core accretion, the relative heavy element to hydrogen ratio in the outer atmosphere may be higher than in the host star. There are two possibilities for this. First, the rock and ice proto-cores erode during accretion of $H_2$-rich gas and subsequent planetary evolution. However, transport and integration of material from eroding cores into the overall planet, may become limited through the metallic and liquid $H_2$/He layers if the planet is large, and if the planet is not fully convective during its entire evolution. The second possibility is inefficient $H_2$ and He accretion, which also causes a higher metal to hydrogen ratio.

(3) The third source of rocky and icy elements in outer planetary atmospheres comes from the late accretion of rocky and icy planetesimals such as asteroidal or cometary bodies. These contributions affect elemental abundances in giant planet irrespective of their formation mode (core accretion or disk instability mechanism). In the solar system, the time of heavy bombardment of the inner planets lasted until about ~3.9 billion years ago, so accretion of larger rocky and icy objects to the giant planets may have proceeded for ~ 0.7 billion years after the start of solar system formation. However, given the mass currently observed in asteroids (less than one Earth mass) and in Kuiper Belt objects (around 1 Earth mass), the absolute amount contributed to giant planets during late-stage evolution may only have been in the Earth-mass range, and may not be a dominant source for most heavy elements in the observable atmospheres of Jupiter and the other outer planets.

## 1.4 Similarities between gas giant planets and brown dwarfs

The densities for the known exoplanets suggest that planets with Jupiter masses and above are $H_2$ and He rich gas giant planets. A first-order assumption for modeling is that such planets have similar elemental compositions as their host stars, that is, near-solar or somewhat enriched in elements heavier than helium. This assumes that there were no physical fractionations of the icy and rocky phases through radial re-distributions so that the formation location of a gas giant planet indeed has the same abundances of the elements as the host star. It further assumes that there were no chemical gas-solid fractionations and no fractionations between the solid icy and rocky phases during planetary accretion or by differentiation process in the planet's interior. A planet's mass and the overall elemental composition are then taken as inputs to model the planet's interior density and temperature structure. The temperature-pressure structure of the outer atmosphere is influenced by the opacity of the gases and condensed phases that are present, and therefore chemistry becomes important. The chemistry is also dependent on T and P, hence the construction of models for the pressure-temperature (P-T) structure of a planet is an iterative process (see e.g., Burrows et al. 2006, Fortney et al. 2006, Hubeny et al. 2003, Marley et al. 2002). The atmospheric P-T structure also depends on the age of the object and its radiative cooling rates. Yet another variable is the distance of a planet to its host star and the radiation that it receives from it.

The outer atmospheric chemistry of most gas giant planets should be dominated by molecular and atomic chemistry so that they resemble either cool methane- T dwarfs, hotter brown dwarfs of type L, or even late M dwarfs. A major factor for the astonishing range in the chemical variations is the different distances of gas giant exoplanets to their host stars (see below). The type of chemistry causes differences in the optical, far-red and near-infrared spectra, and guidelines of what we can expect for exoplanet spectra are already available from the many spectra of ultra-cool stars and brown dwarfs. The appearance and disappearance of certain molecular bands and atomic lines in optical and near IR spectra are applied in the classifications of hot L and cool T dwarfs, which as "brown dwarfs" are the links between real dwarf stars of type M and planets like Jupiter (e.g., Kirkpatrick 2005). The atmospheric chemistry that characterizes the M stars (strong absorption bands of gaseous TiO, VO, CO), L dwarfs (strong alkali lines, strong CO, $H_2O$, FeH, CrH bands), und T dwarfs ($CH_4$, $H_2O$ bands) should be quite similar to what could be seen for exoplanets of different ages and locations from their primaries. The following compares brown dwarf and exoplanet atmospheres.

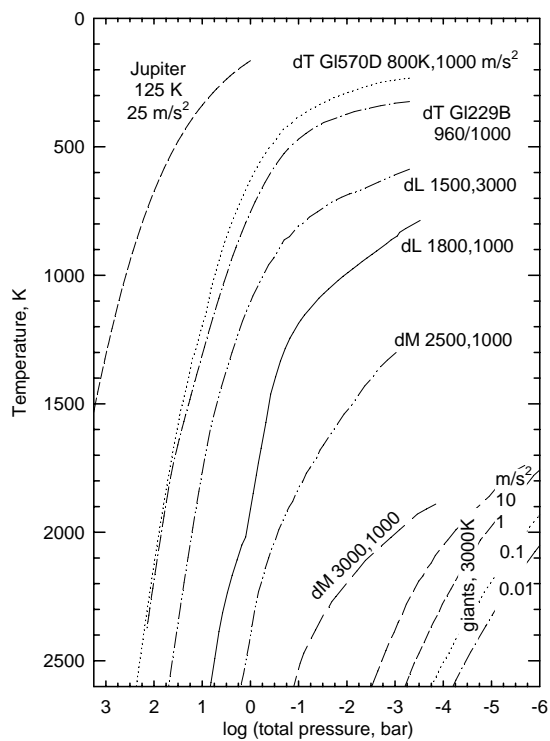

Figure **1.3** Pressure and temperature structures in Jupiter, brown dwarfs, and cool dwarf and giant stars. Numbers after labels indicate effective temperature (K) and surface gravity (in m/s$^2$). Data from Marley et al. 1996, Marley & Saumon 2008, and Marley, personal communication, except for late type stars from Gustafsson et al. 2008.

The P-T structure for Jupiter, two T-dwarfs (Gl570D, Gl229B), two generic L dwarfs, two M dwarfs and M giant stars are shown in Figure 1.3 to give some orientation of the T and P range that we are concerned with (data from Marley et al. 1996, Marley & Saumon 2008, Marley, personal communication). Note that the P and T axes are reversed so that T and P increase towards the origin ("into the object"). The T and L dwarfs have masses above ~13 $M_{jup}$, the minimum mass required for D burning but are below ~ 60-70 $M_{Jup}$ needed for H- burning. The T dwarf Gl229B has an effective temperature $T_{eff}$ ~960 K and a mass of ~ 58 $M_{Jup}$. It is a binary companion to the M dwarf star Gl229A at a separation of ~ 44 AU, so one could think of this brown dwarf as a very massive planet. The T dwarf Gl570D has similar properties $T_{eff}$ = 800 K and 34 $M_{Jup}$ (e.g., Geballe et al. 2001). The P-T models for the T dwarfs have a surface gravity of

~1000 m/s$^2$ (Jupiter's is ~25 m/s$^2$) and the radii of T and L dwarfs are comparable to that of Jupiter and gas giant exoplanets. Models for L dwarfs with relatively similar surface gravities and effective temperatures of 1500 and 1800K are included for comparison. Figure 1.3 shows that objects with higher effective temperatures have higher temperatures at the same total pressure level (e.g., at 1 bar: Jupiter 165 K, Gl229B ~750 K, L dwarf ~1900 K). The age of an object is also an issue because without an internal heat source cooling occurs over time. Thus, objects of similar mass may show different spectral characteristics if they differ in age. The L dwarf model is for an object of ~ 33 M$_{jup}$ and has T$_{eff}$ = 1800 K, whereas the methane T dwarf Gl570D with about the same mass (34 M$_{jup}$) only has T$_{eff}$ = 800 K and different spectral characteristics. Thus, if not externally heated, the effective temperature of objects of similar mass depends on age (see e.g., Burrows et al. 2001, 2003 for P-T evolution of brown dwarfs with time). Since gas giant planets, like brown dwarfs after D-burning, lack strong internal energy sources and are of similar overall elemental composition, the chemistry of their outer atmospheres should change similarly with age, if they are not heated by radiation of their hot stars. Figure 1.4 shows characteristic P-T profiles for Jupiter and Jupiter-mass exoplanets at different separations from their primary star.

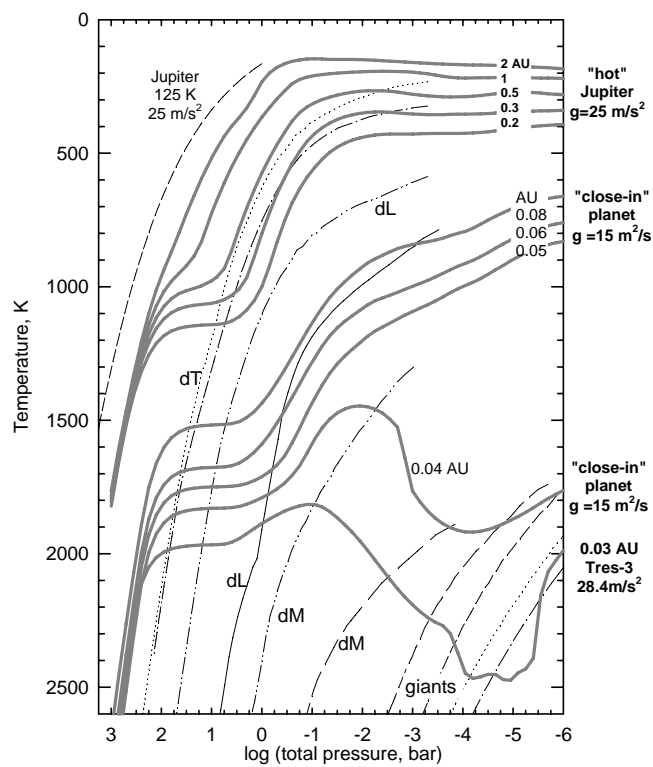

Figure 1.4. Pressure temperature structures of Jupiter-mass planets at different distances (indicated by AU) from their primary stars (planet data kindly provided by J. Fortney; see also Fortney et al. 2008a,b). The P-T structures of some of the more massive objects from the previous Figure are shown with thin lines for comparison.

The model atmospheres of gas giant exoplanets with about the same gravity as Jupiter illustrate how the atmospheric P-T structure changes with a gas giant planet's position from its primary. For comparison, the brown dwarf P-T models from the previous Figure are shown with thin lines. The P-T structures for young (like Jupiter 4.6 Ga ago) planets are shown at the top and plot parallel to the Jupiter profile (Fortney et al. 2008b). These Jupiter-mass planets have an intrinsic $T_{eff}$ of ~100 K if they would not receive any flux from the host star. However, when exposed to radiation from a star like the Sun, the P-T profiles change as shown from top to bottom for planetary orbital distances of 2, 1, 0.5, 0.3 and 0.2 AU, respectively. These P-T profiles become similar to that of T dwarfs, as seen from the coinciding profile of the T dwarf Gl 229B and that of a Jupiter-mass planet at 0.3 AU. However, because Jupiter-mass objects have lower surface gravities than brown dwarfs do, they will have lower pressures in their photospheres if they have the same effective temperatures. Conversely, the depth level from the top of the atmosphere at which the same pressure and temperature is reached in a giant planet is larger than in brown dwarfs because a larger overlying mass is required to build up this pressure. Therefore, there is more opacity and the *effective* temperature of such giant planets will be lower (e.g., compare the T dwarf Gl580D ($T_{eff}$ ~800 K) and the hot Jupiter at 0.5 AU with $T_{eff}$ ~ 370 K.

The second set of profiles in Figure 1. 4 is for planets at smaller orbital distances (0.05 – 0.08 AU). These profiles (provided by J. Fortney) are for a somewhat lower gravity of 15 m/s$^2$, corresponding to Saturn-like planet (~10 m/s$^2$) instead of Jupiter (~25 m/s$^2$). However, the change in gravity does not change the P-T structures as much as does the closer proximity to the primary (taken to be a Sun-like star). The upper parts of these atmospheric P-T profiles resemble those of L-dwarfs as radiative heating from the primary leads to stronger adiabatic expansion of the atmospheres (less dense; shift to lower total pressure) when compared to the planets located at 0.2 AU and beyond. Many known exoplanets orbit their host stars extremely closely (semi-major axis < 0.1 AU) with orbital periods of a few days. The outer atmospheres of these heated "Pegasi-Planets" (after the prototype 51 Peg b), "Roaster Planets" or "Hot-Jupiters" can reach temperatures exceeding 1500 K. Without absorbed flux from the primary, such planets may only have effective temperatures of a few 100 K (depending on age). However, because of the high temperatures, the ongoing chemistry should be comparable to that in hot brown dwarfs (L dwarfs), hence spectra of such exoplanets should be comparable to "free-floating" brown dwarfs of type L. Well studied transiting hot, Pegasi-type exoplanets are TrES-1b and HD 209458 b with outer atmospheres heated to temperatures that correspond to those of brown dwarfs near the L to T dwarf transition ($T_{eff}$ ~1200 to ~1400K).

The two planetary thermal profiles at the highest temperatures are for planets at 0.04 and 0.03 AU, the latter characteristic of the planet Tres-3 b. Exoplanets that are in extremely close orbits with orbital periods of a few hours may develop atmospheric thermal inversions (e.g., Hubeny et al. 2003, Fortney et al. 2008a, Knudson et al. 2009) where high temperatures and tenuous pressures that are comparable to regimes in giant and supergiant star atmospheres; hence the comparison to these much more massive objects here.

## 1.5   Chemistry in gas-giant planets

A useful start of the chemistry discussion is with the more abundant reactive elements (e.g., C,N,O) because their chemistry influences that of less abundant elements. For example, many elements present as monatomic gases can form oxides in cooling gases, depending on the availability of oxygen. In cooler planets, there are several coupled reactions between molecular gases and condensates, but many of these reactions are regulated by the C and O chemistry.

One major chemical difference between T dwarfs and L dwarfs is in their carbon and oxygen chemistry, which is the same as for cool (old and distant) and hot (young/and or heated) Jupiter-mass planets. In T dwarfs and planets like Jupiter, methane is the major C-bearing gas. In L and M dwarfs and close-in gas giant planets, it is CO gas. The chemistry of C and O as functions of T and P are shown in Figures 1.5 and 1.6, which are in the same format as the previous Figures. The curves indicate where the major C and O gases are equal in abundance and define "fields" where a certain gas is dominant. Many details of the C, N, and O distribution and speciation are discussed in Lodders & Fegley (2002). Over a large P-T range, the equilibrium distribution of carbon is controlled by the reaction between CO and $CH_4$ through the net reaction:

$$CO + 3 H_2 = CH_4 + H_2O$$

This reaction, which depends on total P and temperature, describes the equimolar abundance curve of CO and $CH_4$ that bisects the C distribution diagram (Figure 1.5). At higher temperatures and lower total pressures, CO is the major C-bearing gas, and methane abundances gradually decrease when moving away from the equimolar abundance curve into the CO field. For comparison, P-T profiles of Jupiter, and Jupiter-like planets at 1, 0.06 and 0.04 AU from Figure 1.4 are overlaid. The region dominated by CO coincides with the upper atmospheric P-T profiles of hot Pegasi-type planets (and L & M dwarfs). The major C bearing gas in Jupiter's outer observable atmosphere is $CH_4$, and moving Jupiter to 1 AU or 0.2 AU would not change this. By moving a planet closer to the primary, the internal P-T structure shifts closer to the CO=$CH_4$ curve. The CO abundances increase exponentially as one moves more towards the CO field, where CO reaches its maximum abundance. The boundary is crossed if a Jupiter-like planet moves to 0.08 AU or closer, and CO becomes the major gas and $CH_4$ abundances continue to drop towards higher T and lower P.

Objects whose upper atmospheric P-T profiles fall into the $CH_4$-rich field are expected to show methane absorption bands and only little CO absorption, if any at all. However, this only applies if the CO to methane conversion reaches equilibrium values. This is the same problem as noted before for the C chemistry in protoplanetary disks. However, due to the higher total pressures in planetary atmospheres, the CO to methane conversion can proceed down to ~1000 K, whereas in the solar nebula, the temperature for freezing the conversion reaction was about ~1500 K. At quench level temperatures in planetary or brown dwarf atmospheres, the chemical reaction timescale becomes too long when compared to atmospheric mixing timescales, and the CO to methane conversion will not proceed to completion. Thus, the amount of CO is frozen-in or quenched, and CO is overabundant when compared to its expected equilibrium abundance at low temperatures. The overabundance of CO is observed in Jupiter, Saturn, and Neptune and is plausibly explained by quenching of the CO to $CH_4$ conversion due to rapid vertical mixing (see Fegley & Lodders 1994 for a discussion). Given the similarity in chemistry, an overabundance in CO was also predicted for the first discovered methane T dwarf Gl 229b (Fegley & Lodders 1996), which was subsequently observed (Noll et al. 1997). The overabundance of CO is now established for several T dwarfs (e.g., Saumon et al. 2007). Hence quenching of the CO to methane conversion should indeed be an universal phenomenon in atmospheres with thermal profiles in the CO-rich field at higher temperatures and in the $CH_4$ field at lower temperatures. If so, observations and quantization of CO abundances in methane-rich objects provide a tracer of the mixing processes in such objects.

The oxygen distribution is shown in Figure 1.6. The major O bearing gases are CO and $H_2O$. Under conditions where CO is the major C-bearing gas, most O is also in CO but $H_2O$ is always quite close in abundance. With a bulk solar C/O =0.5 and all carbon tied to CO, oxygen should be equally distributed between CO and $H_2O$. However, SiO and other metal oxides reduce the $H_2O$ abundances and the more stable CO becomes the most abundant gas. The conversion reaction of CO to methane produces equimolar amounts of $H_2O$. The $H_2O$ abundance increases proportionally to the decrease in CO. Therefore, the curve where CO and $H_2O$ are equal in abundance is at a similar location as the CO=$CH_4$ abundance curve in Figure 1.5.

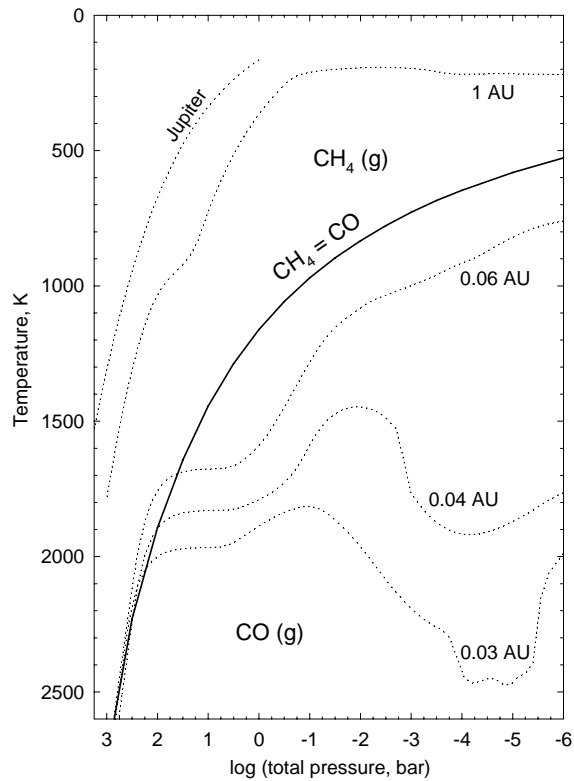

Figure 1.5 Carbon chemistry in low mass objects is dominated by the gases CO (at high T and low P) and $CH_4$ (at low T and high P). The solid line shows where CO and $CH_4$ have equal abundances. The dotted lines are P-T profiles for Jupiter and Jupiter–mass objects at different separations (see previous Figure).

Abundances of O-bearing gases are affected by the formation of condensates in the atmospheres. Forsterite ($Mg_2SiO_4$) and enstatite ($MgSiO_3$) are the two most important sinks for oxygen at high temperatures because Mg and Si are the most abundant rock-forming elements. All high temperature oxides and silicates consume about 20% of total oxygen. The remaining oxygen at lower temperatures is mainly in water vapor, which condenses into a cloud layer around 200-300 K, depending on details of the P-T structure. It is interesting to note that the P-T profile for the hot Jupiter (at 1AU) in Figure 1.6 coincides closely with the condensation curve for water ice (at solar metallicity) over an extended range. Such planets would be prime candidates to show spectral signatures of water ice, whereas in planets like Jupiter, water clouds are buried too deep in the atmosphere to be visible.

The atmospheric P-T profiles of the more strongly heated planets at 0.04 and 0.06AU cross the Mg-silicate stability curves several times. This implies different possible locations within the atmosphere to trap Mg-silicate cloud material from cooler atmospheric regions. However, the atmosphere of the planet at 0.06 AU should be free of Si and Mg-bearing gases at temperatures below ~ 1600 K and an enstatite/forsterite cloud layer is expected between ~1600 to ~2000 K. The planet at 0.04 AU has a more complicated Mg-silicate cloud distribution, and the behavior of condensates in such strongly heated planet with thermal inversions is not fully understood; see e.g., the discussion by Fortney et al. (2008a) on TiO and removal of Ti-bearing condensates.

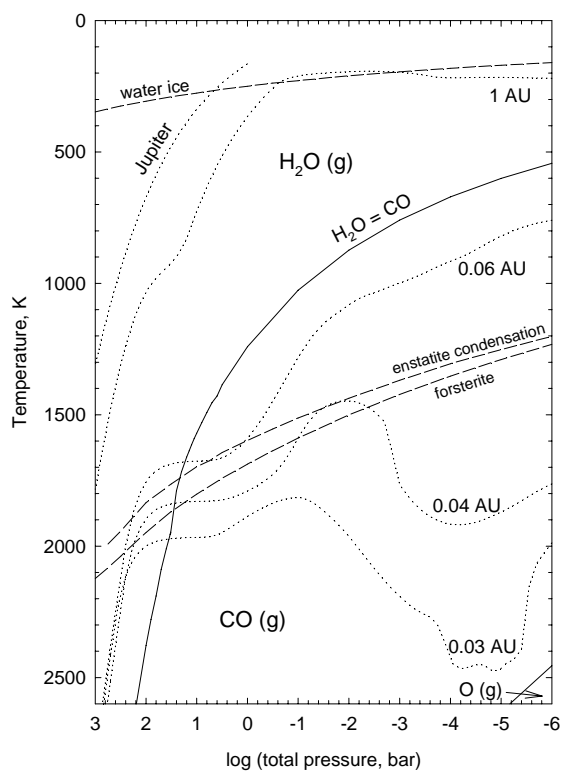

Figure 1.6. Same as Figure 1.5 but for oxygen chemistry.

One can look at all elements in the periodic table and check which gases and condensates are formed at a given temperature and pressure to sort out conditions where important opacity sources (e.g., atomic lines, TiO, VO, CaH, CO, $H_2O$, $CH_4$, FeH, CrH…) appear, what their abundance are, and when these gases are removed by formation of other gases and condensates so that their spectral signatures become lost. This evaluation for all stable elements was done for the Jovian and Saturnian atmospheric pressure and temperature profiles (Fegley & Lodders, 1994). It is an impractical approach to model the chemistry for all of the newly discovered exoplanets this way. However, these results, listed in abbreviated form in Table 2 for the 20 most abundant reactive elements, should be useful to give an indication of possible major gases and condensates. Where appropriate, differences in the major gas abundances expected for hot and cooler objects are indicated.

Table 2. Condensates and gases in solar-composition giant planets

| | Abundance Si=$10^6$ atoms | Condensate Type | Major gases |
|---|---|---|---|
| H | 2.88E+10 | minor fraction in $H_2O$ | $H_2$, ($H_2O$, $CH_4$) |
| O | 15500000 | oxides, silicates, water | $H_2O$; CO high T |
| C | 7080000 | $CH_4$ | $CH_4$ low T, CO high T |
| N | 2090000 | $NH_4SH$, $NH_3$ | $NH_3$ low T, $N_2$ high T |
| Mg | 1020000 | $Mg_2SiO_4$, $MgSiO_3$ | $Mg(OH)_2$, MgOH, MgH, Mg |
| Si | 1000000 | $Mg_2SiO_4$, $MgSiO_3$ | $SiH_4$, SiO, SiS |
| Fe | 838000 | Fe-alloy | Fe, $Fe(OH)_2$, FeH |
| S | 445000 | $NH_4SH$; minor MnS, ZnS, $Na_2S$ | H2S |
| Al | 84100 | $Al_2O_3$, Ca-Al-oxides | $Al_2O$, AlH, AlOH, $HAlO_2$ |
| Ca | 62900 | Ca-Al-oxides, Ca-titanates | Ca, $Ca(OH)_2$, CaOH, CaH, CaS |
| Na | 57500 | $Na_2S$ | Na, NaH, NaCl, NaOH |
| Ni | 47800 | NiFe-alloy | Ni, NiH, NiS |
| Cr | 12900 | Cr-metal, $Cr_2O_3$ | Cr, CrH, CrS, CrO |
| Mn | 9170 | MnS | Mn, MnH, MnS |
| P | 8060 | $NH_4H_2(PO_4)_3$, minor $Cu_3P$ | $PH_3$, $PH_2$, $P_2$, PO |
| Cl | 5240 | NaCl, KCl, $NH_4Cl$ | HCl, NaCl, KCl |
| K | 3690 | KCl | K, KCl, KOH |
| Ti | 2420 | Ca-titanates, e.g. $CaTiO_3$ | TiO, $TiO_2$, Ti |
| Co | 2320 | FeNi alloy | Co, CoH |
| Zn | 1230 | ZnS | Zn, ZnH, ZnS |
| F | 841 | NaF, KF, NH4F | HF |
| Cu | 527 | $Cu_3P$ | Cu, CuH |
| V | 288 | V-oxides into Ca-Ti & Ca-Al-oxides | VO, $VO_2$, V |
| Ge | 121 | Ge, minor GeTe | $GeH_4$, GeS, GeSe, GeTe |
| Se | 65.8 | PbSe, other selenides | $H_2Se$ |
| Li | 55.5 | LiF, $Li_2S$ | Li, LiF, LiCl, LiOH, LiH |
| Ga | 36.0 | GaS | GaOH, $Ga_2S$, GaS |
| Sc | 34.2 | $Sc_2O_3$ | ScO |
| Sr | 23.6 | SrS | $Sr(OH)_2$, SrOH, Sr |
| B | 17.3 | $H_3BO_3$ | $H_3BO_3$, $NaBO_2$, $KBO_2$, $HBO_2$ |
| Br | 11.3 | NaBr, KBr, RbBr, $NH_4Br$ | HBr, alkali bromides |
| Zr | 11.2 | $ZrO_2$ | ZrO, $ZrO_2$, ZrS |
| Rb | 6.57 | RbCl, RbBr | Rb, RbF, RbCl |

## 1.5.1 Condensate Clouds

Thermodynamic equilibrium and kinetic calculations are useful to find which elements and chemical species are present as functions of T, P, and metallicity. The computation of gas-phase equilibria is straight forward, but condensate formation in gas giant planets must consider that initially-formed condensates sediment into cloud layers and do not react with gases at higher (cooler) altitudes. The chemical condensate treatment requires the distinction of two possible end-member settings: Condensate formation in a strongly gravitationally bound atmosphere and condensate formation in a low gravity environment such as in protoplanetary disks. Figure 1.7 illustrates the differences in condensates for some elements. In a bound atmosphere, condensates forming directly from the gas at high temperatures ('primary condensates') settle due to the influence of gravity and form cloud layers. These primary condensates do not react with the cooler gas at altitudes above the condensate clouds so they are out of equilibrium with the gas in the overlying atmosphere. In contrast, condensates in protoplanetary disks can remain dispersed in the gas and react with it to form secondary condensates during cooling. For example, in both cases, Fe-metal is the first Fe-bearing condensate. However, in planetary atmospheres, $Fe_3P$ and FeS cannot form because Fe is in a deep cloud. Then P condenses mainly into $NH_4H_2PO_4$, and S condenses into sulfides such as $Na_2S$, MnS, and mainly into $NH_4SH$.

The condensate cloud layer concept (often called rainout) was initially developed for the Jovian planets (Lewis 1969, Barshay & Lewis 1978), and it has been found that only models including cloud sedimentation are capable of matching observed brown dwarf color diagrams and spectra. Hence what works for Jupiter and brown dwarfs should work for exoplanets with masses in-between.

The cloud masses that will be available depend on the availability of the constituent elements in the condensates. Table 2 shows that the most abundant rock-forming elements Mg, Si, and Fe produce massive clouds of forsterite, and enstatite, and liquid iron, respectively. The other less abundant rock-forming elements such as Ca, Al, Ti, Mn, Cr, and the alkalis will produce less-massive clouds. In addition to the more refractory (high thermal stability) condensates, O, S, N, and C form abundant condensates at temperatures below ~200 K. Giant planets with low-temperature upper atmospheres can have massive icy cloud layers of water (ice or liquid), $NH_4SH$ (s), $NH_3$ ice and $CH_4$ ice. However, they will also have the full sequence of deeper seated cloud layers containing the more refractory elements in the interior.

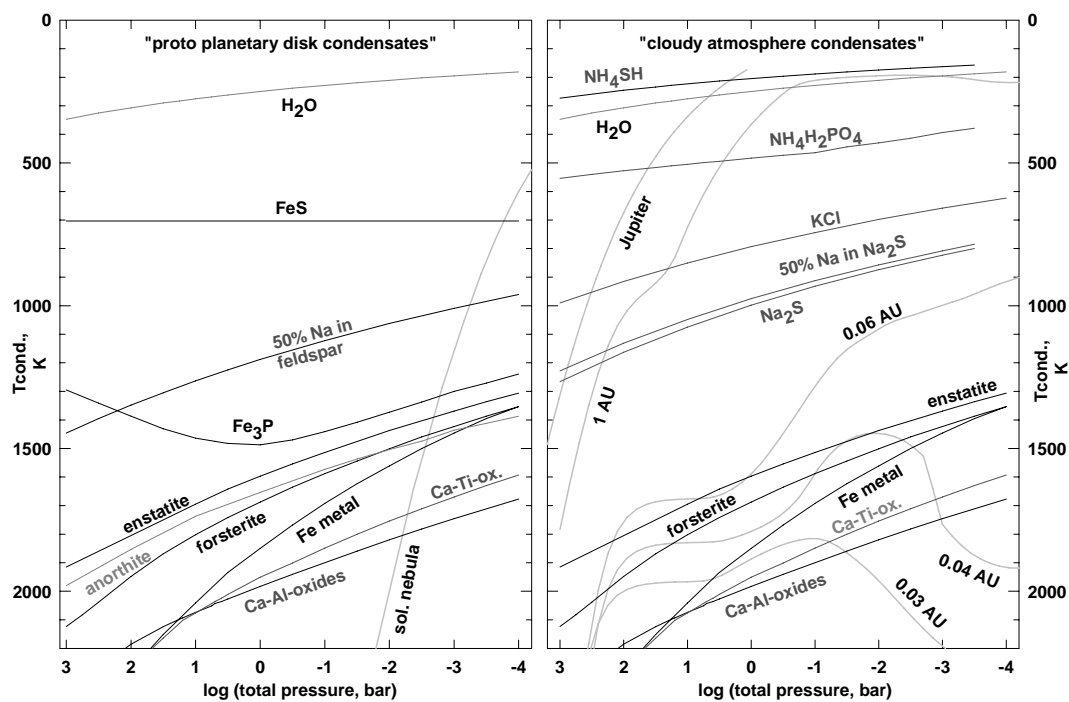

**Figure 1.7** The left graph shows some of the condensates expected in protoplanetary disks. There primary, high T condensates (e.g., Fe metal) can continue to react with gas to form secondary condensates (e.g., FeS) at lower temperatures. The right side graph for planetary atmospheres shows condensates that appear if primary condensates settle into cloud layers. There are no secondary Fe-bearing condensates and different types of condensates appear at low T. Note that high T and high P are towards the bottom left in the graphs.

Once a condensate becomes stable, the abundance of the constituent gases exponentially decreases with decreasing temperature in the atmosphere. With the removal of the gases, their opacity also disappears. Therefore, the cloud layer structure in the atmosphere becomes traceable through the changes in gas chemistry. The expected spectral features caused by removal of gases into condensate clouds in giant planets should be similar to that known for the hot to cool brown dwarf sequence (see the discussion in Lodders & Fegley 2006).

Extremely heated gas-giant planets will have very strong TiO and VO bands in the optical, and $H_2O$, CO and FeH bands in the near-infrared. Near ~1800-2200 K, refractory ceramics such as corundum ($Al_2O_3$) or Ca-aluminates (such as hibonite, $CaAl_{12}O_{19}$) and Ca-titanates (such as perovskite, $CaTiO_3$) become stable. This leads to a gradual disappearance of TiO and VO bands since VO follows TiO by forming solid solutions with the Ti-bearing condensates. Similarly, Ca and Al-bearing gases are removed. The exact nature of the refractory condensate is sensitive to the total pressure (e.g., see Lodders 2002). The first Ca, Al condensate at high total pressures in the deeper interiors of cool Jupiter like giant planets is melilite (a solid solution of gehlenite $Ca_2Al_2SiO_7$ and akermanite $Ca_2MgSi_2O_7$). However, it is hibonite below ~1 bar, and $Al_2O_3$ below 0.01 bar, which is more relevant to Pegasi-type planets or extremely heated exoplanets. The initial Al bearing condensate in Pegasi planets is Ca-poor, which affects Ca gas abundances. The Ca (and Al) chemistry determines which Ca-titanate forms, which in turn is important for the TiO gas abundances.

Since the temperature of gas giant planets in close orbits is regulated by the irradiation from their host stars, one can expect that the closest planets (around the same type of star) are the most likely ones to be heated above temperatures necessary for the condensation of ceramics. In that case, TiO and VO remain and their opacity determines whether a Pegasi type planet can develop a stratospheric temperature inversion (e.g., Huebeny et al. 2003, Burrows et al. 2007, Fortney et al., 2008a). The photometric data and spectra of HD 209458b (Knutson et al. 2008, Desert et al., 2008) seem to be consistent with the presence of TiO and VO and a low pressure inverted temperature structure in this exoplanet.

Molecular bands of CrH and FeH become stronger in cooler atmospheres but decrease in strength when Fe-metal and, depending on total pressure, Cr or $Cr_2O_3$ condensation occurs. This can remove all Fe and Cr gases, respectively, around ~1200-1500 K. The removal of Fe into a cloud layer is also the reason why $H_2S$ remains in the atmospheric gas at lower temperatures. If Fe cloud settling did not occur, $H_2S$ would be completely absent from the upper atmospheres because formation of FeS (troilite) from Fe metal grains with $H_2S$ gas starting at ~700 K consumes all $H_2S$ gas (solar Fe/S~2). However, formation of secondary FeS and removal of $H_2S$ at 600-700K is at odds with the *Galileo* probe observations on Jupiter. The *Galileo* entry probe mass spectrometer (GPMS) detected $H_2S$ at ~3 times the solar $S/H_2$ ratio in Jupiter's atmosphere (Niemann *et al*. 1998) which is only possible if $H_2S$ is not removed. Because $H_2S$ is observed in the Jovian and Saturnian tropospheres at altitudes below the $NH_4SH$ cloud condensation level, the condensate formation models must call for the depletion of Fe metal into a cloud deep in the Jovian and Saturnian atmospheres.

In the temperature range of Fe-metal condensation, Mg-silicate (forsterite and enstatite) cloud formation is expected as well. This will remove all Mg (e.g., Mg, MgH, MgOH) and Si (e.g., Si, SiO, SiS, $SiH_4$) gases from the cooler atmosphere. As for sulfur, Jupiter and Saturn are test cases for the cloud layer approach. The absence of silane ($SiH_4$) and the presence of germane ($GeH_4$) in Jupiter and Saturn is due to depletion of refractory Si, but not of volatile Ge, by condensate formation deep in their atmospheres (Fegley & Lodders 1994). Silicon is much more abundant than Ge in a solar composition gas (atomic Si/Ge ~8300) but there are only observational upper limits for $SiH_4/H_2 \sim 1\times10^{-9}$ (1 ppb) on Jupiter and Saturn. For comparison, the protosolar $Si/H_2$ molar ratio is $8.23\times10^{-5}$; ~82,000 times larger than the observational upper limit on the silane abundance. In contrast, the observed $GeH_4/H_2$ is ~0.7 ppb on Jupiter and ~0.4 ppb on Saturn (Lodders & Fegley 1998). These values are closer to the protosolar $Ge/H_2$ molar ratio of 9.9 ppb and the difference arises because not all Ge in Jupiter and Saturn is present as $GeH_4$ (see Fegley & Lodders 1994). The presence of Mg-silicate clouds should be testable by searching for $SiH_4$ (which has strong IR bands) in the atmospheres of methane rich planets, or for SiO and SiS in hot CO rich planets. These gases should be absent or depleted above the $Mg_2SiO_4$ and $MgSiO_3$ clouds if the cloud condensation models are correct.

Enstatite is the lowest temperature condensate of the major element (Al, Ca, Ti, Mg, Si, Fe, Ni) condensates. Thus, all of the major rock-forming elements are out of the gas above the enstatite cloud. However, lines of monatomic alkali metals K, Na, Cs and Rb should remain prominent and persist to lower temperatures that correspond to hotter methane-rich objects (Figure 1.5). Monatomic K gas is observed in T dwarfs such as Gl 229B and Gl 570D (e.g., Burrows *et al.* 2000, Geballe *et al.* 2001). Monatomic Na, at less than the solar Na abundance, is inferred for the exoplanet HD 209458b (Charbonneau *et al.* 2002). The monatomic gases are dominant until they gradually convert to NaCl, KCl and other gases such as oxides, hydroxide, and hydrides, before being removed as sulfide and halide condensates. Sodium is much less abundant than S, so all Na can be sequestered into a $Na_2S$ cloud but only 6.5% of all sulfur would be lost, which does not lead to a significant loss in total $H_2S$.

The presence of Na and K in cooler objects gas requires that refractory rock-forming elements such as Al, Ca, and Si are depleted by condensate cloud formation deep in the atmosphere. Otherwise, Na and K-gases would condense into silicate minerals such as $XAlSi_3O_8$ (albite (X=Na) and orthoclase (X=K)) at high temperatures, and Na and K gas would be depleted from the observable atmosphere. However, if Al, Ca, and Si are in deep condensate clouds, $XAlSi_3O_8$ condensates cannot form and Na and K stay in the gas. Only model spectra that consider Al, Ca, and Si removal into deep cloud layers show good agreement with the higher observed K I line strengths for Gl 570D and Gl 229B (e.g., Geballe *et al.* 2001). The chemistry of Li and the other alkalis Rb, and Cs is similar. Instead of being removed from the gas by reaction with high-temperature feldspar, they only form their own sulfide (e.g., $Na_2S$) and halide condensates at temperatures 500 K lower or more than e.g., Mg-silicates.

The switch from CO to methane should cause an onset of absorption in the 1.6, 2.2 and 3.3 μm methane bands; similar to what was expected and found in the L-dwarf spectral series (Noll et al. 2000) With methane becoming more abundant, the infrared colors will change because of the strong IR methane absorption bands, as is known for the change in L and T dwarfs. In cool methane dominated objects, the alkali lines have vanished. Methane, water and ammonia characterize the cooler objects until condensation of liquid water or water ice and solid $NH_3$ leaves only methane behind in an otherwise He and $H_2$-rich atmosphere. Removal of methane from the atmosphere into clouds requires very low temperatures (< 50 K).

## 1.5.2 The effects of varying the C/O ratio on gas giant planet chemistry

### 1.5.2.1 Gas chemistry variations through changes in C/O ratio

Variations in the C/O ratio affect the chemical composition of the atmosphere because the CO, methane and water equilibria are changed. This affects the interior evolution and spectra of gas giant planets. For example, Fortney et al. (2005, 2006) show that changes in metallicity and C/O ratios from solar values strongly influence the emergent spectra of brown dwarfs and hot exoplanets. Changes in C/O mainly affect abundances of gases containing C and/or O. To a first approximation, abundances of gases such as $NH_3$ and $N_2$ are not affected much because the most abundant N-bearing gases contain neither C nor O. However, the increasing abundances of gases such as HCN that contain C and are of minor importance in a solar composition gas may increase quite strongly.

Figure 1.8 shows the gas abundances as a function of C/O ratio at 1100 K and 0.01 bar to illustrate trends. At this temperature, C is mainly distributed between CO and $CH_4$, and O between CO and $H_2O$ in a solar composition gas. The gas abundances are calculated for C/O ratios from solar to unity by increasing the C-abundance (closed symbols in Fig. 1.8) or by decreasing O from the system (open symbols). The partial pressures of CO, $CH_4$, $C_2H_2$, HCN, and CN increase whereas that of $H_2O$ declines over orders of magnitude at otherwise constant conditions. The increase in the CO fugacity is relatively modest (a factor of 2), which reflects increasing the C/O from 0.5 to unity and that most C is in the very stable CO molecule. Therefore, changes in CO opacity, important for the 4.5 micron spectral window, decreases. A particularly strong increase is for $C_2H_2$ and HCN which can become new opacity sources in the infrared. On the other hand, the $H_2O$ fugacity drops by several orders of magnitude if the C/O ratio is increased from solar to unity because with a relatively high C content, more O is in CO and less O is available for water and other O-bearing molecules.

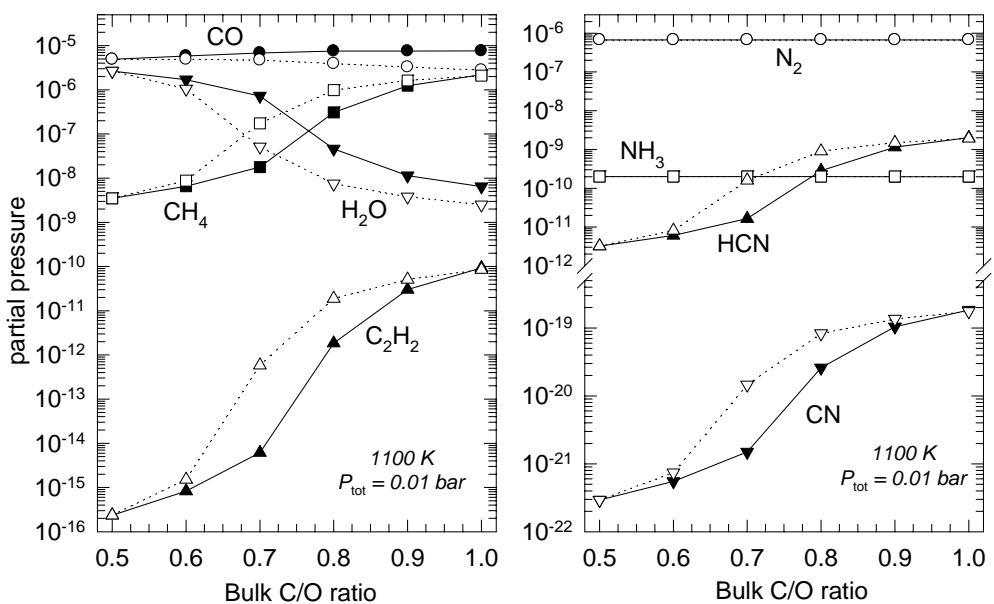

Figure 1.8. Variation of gas chemistry of carbon and oxygen (left) and nitrogen (right) as a function of C/O ratio at 1100 K and 0.01 bar. The C/O ratio was increased from solar by increasing the total C abundance (closed symbols) or decreasing the total oxygen (open symbols). Note the break in the concentration scale for nitrogen gases.

Figure 1.8 also illustrates that it matters how the C/O ratio is altered. A decrease in C/O ratio by O removal yields a larger effect than adding C because the absolute amount of O-bearing molecules decreases. This is the reason why the $H_2O$ abundance decreases much more strongly when O is removed. Since water is a major opacity source in gas giant exoplanets, substantial decreases in water abundances should change the spectra in the 8 micron region of these objects (see e.g., Fortney et al. 2006). Together with observations of CO, this may constrain planetary C/O ratios, which has possible implications for planetary formation scenarios. If a gas giant planet has a different C/O ratio from its host star, the variation of the C/O ratio must have to do with the planet formation itself, if chemical fractionations (such as water ice condensation) in the planet can be ruled out.

### 1.5.2.2 Possible scenarios to alter C/O ratios during planetary formation

There are two possibilities to change the C/O ratio in a giant planet from the value of ~0.5 in an otherwise solar composition gas. For an increase in C/O in an accretion disk, the plausible mechanisms are to increase the carbon abundance or decrease the oxygen abundance. For a decrease in C/O, an increase in O is the most plausible possibility.

Assuming that gas giant planets form by core accretion, an increase in the C abundance occurs if abundant interstellar or locally produced organics are incorporated into the rocky proto-core of a giant planet. As discussed earlier, tarry organics could exist in larger amounts in planetary accretion disks beyond an organic condensation/evaporation front (a tarline). This hypothesis is analogous to assuming that planets like Jupiter grew from a protocore composed of rock and water ice, at an orbital distance near or beyond the water ice condensation/evaporation front (the snowline). Organic substances may be similarly suitable to increase the mass density of the disk to aid rapid formation of a core and by "gluing" rocky and carbonaceous substances (see, Lodders 2004).

The compositional outcome of these two cases is quite different. If a planet forms with excess water ice, the resulting C/O ratio is lower than the host star's C/O ratio. On the other hand, accretion and growth from a carbonaceous and rocky core leads to a C/O ratio above the host star's ratio.

Organics constituting a tarline require higher temperatures to evaporate than water ice, hence there can be regions in accretion disks that are rich in carbonaceous material, but poor in water ice. The stability of the organics would be enhanced if water was cold-trapped in a more distant region from the central star, because then water vapor, which is a potential oxidant for the carbonaceous material, is also less abundant.

The cold-trapping of water ice also provides a possibility to rise the C/O ratio because it is removes oxygen in the planet forming region. Radial diffusive redistribution of gas and water cold-trapping at and beyond the snow-line would dehumidify the region inward of the snowline (e.g., Stevenson & Lunine 1988). This leads to increased C/O ratios between the star and the snowline. Even if half of all O is in CO as expected from the solar C/O = 0.5, removal of other oxygen moves the C/O ratio toward unity. If planets then acquire water depleted gas, they may have a higher C/O ratio. However, if gas and water ice are not physically fractionated before accretion to a planet, the *bulk* C/O ratio remains solar.

Thus, a planet with a higher C/O ratio than its host star formed either in a water depleted region or in a region that was enriched in carbonaceous material. Since fast growth of a massive rocky proto-core may be required for making gas giant planets, the first option, accretion in a water-depleted region could cause a problem because the surface mass density of condensed rocky mass alone is too low for fast growth. In this case, the presence of abundant organic materials could help to increase the mass density. This limits the origin of proto-core formation for gas giant planets with higher C/O ratios to regions in protoplanetary disks below ~400 K, where carbonaceous material remains stable.

The other extreme is a gas giant planet with a C/O ratio smaller than in its primary which formed by preferential accretion of water-ice. Water ice trapping from the inner planetary system must lead to ice pile-up at the snowline and beyond, which would be an ideal source. Such a pile-up of water is probably recorded by the water-rich planets Uranus and Neptune. If these planets formed near or beyond the snowline with abundant water ice, water enhancements and a smaller C/O ratio would result (e.g., Lodders & Fegley 1994). This formation scenario applies to disk regions with temperatures below ~ 180 K to stabilize water ice.

## 1.6 Outlook

The chemistry of exoplanets is somewhat difficult to talk about since chemical information is mainly available in indirect form through density measurements, and direct spectroscopic observations so far provide somewhat limited results. This will undoubtedly change in the coming years. Most known exoplanets are gas-giant planets, and most modeling is thus available for gas giant exoplanets. This will also undoubtedly change in the coming years, when the upcoming missions begin to discover terrestrial-like planets. However, even without too many observations on chemistry yet, exoplanet chemistry is predictable. It is constrained by the abundances of the elements and the physical conditions in planet formation environments as well as in the planet itself. The observable chemistry of brown dwarfs, the exoplanets' larger "cousins", also has provided valuable guidelines to what can be expected in gas giant exoplanet spectra, and the known chemistry of the terrestrial planets in the solar system will be equally valuable in interpreting the observations of exoplanets to come.

Acknowledgments: I thank Jonathan Fortney and Mark Marley for kindly providing model atmospheres, Bruce Fegley for discussions, and Rory Barnes for patiently waiting for the manuscript. This work was supported in part by NSF-grant AST 08-07377 and NASA grant NNG06GC26G.